RESEARCH-ARTICLE

# Usability Study of Security Features in Programmable Logic Controllers

**KAREN LI**, University of Bristol, Bristol, U.K.

**KOPO MARVIN RAMOKAPANE**, University of Bristol, Bristol, U.K.

**AWAIS RASHID**, University of Bristol, Bristol, U.K.

# Usability Study of Security Features in Programmable Logic Controllers

Karen Li
qc19463@bristol.ac.uk
University of Bristol
Bristol, United Kingdom

Kopo M. Ramokapane
University of Bristol
Bristol, United Kingdom
marvin.ramokapane@bristol.ac.uk

Awais Rashid
University of Bristol
Bristol, United Kingdom

## ABSTRACT

Programmable Logic Controllers (PLCs) drive industrial processes critical to society, for example, water treatment and distribution, electricity and fuel networks. Search engines, e.g., Shodan have highlighted that PLCs are often left exposed to the Internet, one of the main reasons being the misconfigurations of security settings. This leads to the question – why do these misconfigurations occur and, specifically, whether usability of security controls plays a part? To date, the usability of configuring PLC security mechanisms has not been studied. We present the first investigation through a task-based study and subsequent semi-structured interviews (N=19). We explore the usability of PLC connection configurations and two key security mechanisms (i.e., access levels and user administration). We find that the use of unfamiliar labels, layouts and misleading terminology exacerbates an already complex process of configuring security mechanisms. Our results uncover various (mis-) perceptions about the security controls and how design constraints, e.g., safety and lack of regular updates due to long term nature of such systems, provide significant challenges to realization of modern HCI and usability principles. Based on these findings, we provide design recommendations to bring usable security in industrial settings at par with its IT counterpart.

Authors' Contact Information: Karen Li, qc19463@bristol.ac.uk, University of Bristol, Bristol, United Kingdom; Kopo M. Ramokapane, University of Bristol, Bristol, United Kingdom, marvin.ramokapane@bristol.ac.uk; Awais Rashid, University of Bristol, Bristol, United Kingdom.

## 1 INTRODUCTION

Industrial Control Systems (ICS) manage industrial processes for critical infrastructures like electricity and water [3, 18, 37]. Programmable Logic Controllers (PLCs), essential to ICS, control physical processes via sensors and actuators, and are configured through Human Machine Interfaces (HMIs), which present data to operators. High-profile attacks, such as on the Ukrainian power grid [1] and the Florida water treatment plant [2], highlight ICS security risks. These are worsened by PLCs' internet visibility, often due to misconfigurations, with many PLCs accessible via Shodan [3]. Shodan is a search engine that indexes internet-connected devices, including ICS and PLCs. The accessibility of PLCs through Shodan poses significant risks, as attackers can locate vulnerable PLCs and exploit security flaws, potentially disrupting industrial processes and causing physical damage. Foley's year-long study [14] on a misconfigured PLC in a public utility revealed ongoing security lapses, with a critical port left open despite corrections. Such cases, coupled with the widespread online presence of PLCs, underscore the need to explore why these misconfigurations occur and the role of usability in security. Efforts to protect PLCs include studies on attacks [36, 63], intrusion detection [16, 19, 24, 52], and testbeds [15, 24]. However, the usability of security mechanisms for configuring PLCs remains unexamined. This paper addresses this gap by investigating usability challenges faced by operators during PLC configuration, affecting security.

In IT environments, usability in security configurations is well-studied [4, 20, 25, 29, 34, 55, 56, 61]. Our study examines PLC security configuration usability, defined by operators' ability to accurately and efficiently configure security within resource constraints. We conducted an exploratory study with 19 participants, evaluating connection configurations and common PLC security mechanisms using a Siemens S7-1200 [4]. The paper is organized as follows: Background provides an overall view on IT and ICS differences and how usability is less understood in ICS. Methodology provides a detailed experimental design to how selected our participants and done our data analysis. Section 3 describes our results reveal complexities in configuring PLC security due to complex navigation, misleading terminology, and unfamiliar icons. Discussion discusses the implications of these findings, including the lack of product usability and experience vs usability. Finally, we end with the conclusions and recommendations for future work.

In summary, we contribute:

[1] https://www.bbc.co.uk/news/technology-38573074
[2] https://bit.ly/3uZK7zH
[3] https://www.shodan.io
[4] Siemens boasts the highest global share of the PLC market [45].

- The first study on PLC security configuration usability, highlighting commonalities across Siemens, Allen Bradley/Rockwell, and Mitsubishi PLCs, and identifying usability challenges through task flow diagrams.
- Identification of factors underlying usability challenges, such as PLC design considerations and HMI limitations, requiring more than interface updates or adding capacitive screens.
- Emphasis on understanding PLC operators' roles to improve ICS security, showing that PLC procedures prioritize functionality and reliability over ease of use, a key theme in usable security studies.

# 2 BACKGROUND

In this section, we provide background on Industrial Control Systems (ICS) and Information Technology (IT) environments, emphasizing their differences. We revisit the principles of usable security, highlighting the disparities between ICS and IT systems in this context. Our goal is to illuminate the challenges within ICS environments that require attention. IT and ICS environments differ significantly, necessitating tailored security approaches. ICS equipment is designed for harsh environments and runs on legacy firmware, with updates being rare and requiring extensive predeployment testing [51]. In contrast, IT systems are modern, powerful, and frequently updated. ICS interfaces, often small and outdated, rely on HMIs, whereas IT systems have more intuitive and regularly updated user interfaces. Safety takes precedence in ICS, sometimes resulting in shorter passwords [38]. Security in ICS is still developing, with research focusing more on functionality and vulnerabilities than on usability and security updates. Studies reveal barriers in establishing robust security cultures, inadequate inter-function communication, and a lack of ICS cyber security expertise [10]. A deeper understanding of the ICS environment, devices, and personnel is crucial. The ISO 9241-11:2018 standard defines usability as the extent to which users can achieve specific goals with effectiveness, efficiency, and satisfaction [22]. Usable security integrates human-centered design with strong security measures, aiming to create systems that prioritize user experience without compromising security [46]. Research in this field has led to a better understanding of user behaviors, mental models, and decision-making processes [1, 23, 43]. It aims to design intuitive, efficient, and effective systems that also provide necessary security features, reducing user errors in security-related tasks [48, 60]. In IT environments, extensive research has resulted in mature principles and design patterns for configuring security mechanisms. This has led to more user-friendly security measures, such as biometric authentication and password managers. Research has also improved the communication of security risks and warnings in web browsers and influenced policies, educational initiatives, and standards [2, 9, 11, 28, 44, 53, 62].

PLCs are specialized digital computers used in industrial automation, known for their flexibility in programming, reliability in harsh environments, real-time operation capabilities, modular design, and integration with various industrial equipment [6]. They play a pivotal role in ICS by automating processes, ensuring sequence control, diagnosing faults, logging data, and interfacing with other control systems. Another study[42] highlighted PLC security issues in ICS and introduced the code security, firmware security, network attack and some other techniques for PLCs. But, usability in ICS is less understood, with many unknowns about the constraints and challenges faced by operators and engineers. Prior works [30, 31]laid the groundwork for a usable security model for ICS, highlighting the importance of addressing poor usability. Another study suggested implementing tangible interfaces to improve usable security in ICS[32]. These studies indicate that potential attacks such as tampering and spoofing are not only due to the absence of security mechanisms but the lack of consideration in usable security during requirement stage. For example, a complicated login procedures, though increases security from attackers, it can also lead operators to seek shortcuts such as saving passwords in insecure ways or ignoring warnings. This is one of the security usability trade off as attackers exploit this by creating a simpler, convincing fake login page called phishing. Users would seek this and unknowingly provide their credentials to attackers. These would have dire implications such as safety - usability directly influences the safety and reliability of these industrial operations making intuitive interfaces critical. This paper provides empirical evidence of usability issues in ICS environments.

# 3 METHODOLOGY

To assess the usability of PLC security mechanisms, we conducted a qualitative study involving 19 participants from October 2019 to March 2020. Participants were tasked with configuring the security mechanisms of a Siemens S7-1200 PLC using a cognitive walkthrough and the think-aloud approach [5, 40, 41, 50]. This method required participants to articulate their decision-making process during each action while our team observed and took notes. Our primary objective was to explore the usability challenges encountered by operators while configuring PLC security mechanisms. Rather than focusing on specific usability metrics, our approach aimed to capture nuanced user experience, which is often multifaceted and not solely quantifiable. Users' emotions, satisfaction, and overall experience often necessitate an in-depth and qualitative exploration [8, 33].

## 3.1 Ethical considerations

This study was reviewed and approved by our institutional ethics review board (IRB). Before taking part in the study, participants were given a participants information sheet and a consent form. The participants information sheet contained all the information about the study, the study's main objectives, what was expected of the participants, and how data from the study will be handled and stored.

## 3.2 Experiment Design

We designed a task-based study focusing on configuring a Siemens S7-1200 PLC followed by a semi-structured interview. Siemens PLC mechanisms are configured through the *Totally Integrated Automation Portal* [5] (TIA portal). TIA is designed to give operators unrestricted access to many Siemens digitized automation services.

*Task-based exercises.* Our task-based exercises aimed to observe participants configuring the mechanisms of a PLC. They were not

[5]https://new.siemens.com/global/en/products/automation/industry-software/automation-software/tia-portal.html

**Table 1: ICS and IT environment differences according to Stouffer et al.'s NIST guide (800-82) to ICS security [51]. Detailed differences are in the guideline.**

| Category | IT Environment | ICS Environment |
|---|---|---|
| Operating system | Managed by IT professionals. | Managed by Control Engineers and Operators, often lacking default security capabilities. |
| Change Management | Software updates with automated deployment tools. | Infrequent software changes requiring vendor testing and scheduling for outages. |
| Communication | Standard IT networking practices. | Proprietary and standard protocols requiring control engineer expertise. |
| Component Location | Usually local and accessible. | Often isolated, remote, and requiring significant physical effort to access. |
| Resource Constraints | Designed to accommodate additional applications, including security solutions. | Limited computing resources hindering additional security capabilities. |

**Task 1**

*The purpose of this task was to investigate the usability of the process of connecting a PLC to a PC. We studied this procedure because connecting a PLC to PC is a standard procedure in ICS settings for any configurations. To investigate this, we asked participants to link a provided PLC to the PC that was already connected physically to a PC through an ethernet cable. We expected participants to link and verify the connection between the PLC and the PC, either using the IP Address we provided or the Ethernet cable. A successful connection is achieved when the participant can flash the LED lights on the PLC.*

To connect a Siemens S7-1200 PLC to a computer, a user would need to load a TIA portal. If there is no existing project, then they would need to create a new one by clicking on "create new project." After a project has been created, the user then needs to configure the device. To configure the device, they first need to add the device, which involves finding the right PLC type and version from the provided dropdown menus.

- The next step involves them to right-click on the added PLC and then properties.
- On the resulting screen, under the "General" tab, the participant is then required to locate and click on 'ethernet addresses' on the inside left pane.
- The user is then asked to provide the method of connection, ethernet cable, or provide a valid IP address.
- After this, the user must then click "Go online". To confirm if the PLC is connected, the participant must search for the PLC by clicking on "start search".
- Once the PLC is found, the option to flash the LED lights on a PLC is made available. By checking the "Flash LED" box will flash the LEDs on the actual PLC device. This is optional.
- Once completed, the user would click "Go Online". The "Go Online" means connecting to the PLC so one can upload logic or send commands.

**Task 2**

*The purpose of Task 2 was to examine the usability of configuring access level control mechanisms. Access level control is an important security mechanism for managing or restricting access to different features (e.g., memory) and configurations of a PLC. To explore this, participants were asked to provide "Complete protection" to all access levels (i.e., restrict access without password) of the PLC. Participants were expected to navigate to the access level control menu, select the appropriate level, and then provide passwords to all the other levels. A successful task involves a participant being able to navigate to access level control mechanisms, selecting "Complete protection" option, and then setting passwords to all the other levels.*

Under Siemens S7-1200, users can configure four access level. Setting the access level and password restricts access to some functions and memory areas without a correct password. The four levels are as follows:
***No protection*** - Configurations can be read and modified by anyone;
***Write protection*** - Configurations can be read without a password while every other modification require valid password;
***Write/Read protection*** – Gives access to HMI and diagnostic data only;
***Complete protection*** – No access and modification to configurations without valid password.

- To configure these access levels, the user must first right click on the Central Processing Unit (CPU) in the project and then on properties on the resulting panel (From Task 1).
- Then, under the "General" tab, they are required to click on "Protection & Security" and then click "Access Level," which will display all the access level controls.
- To configure complete protection, which means setting passwords to all the other three levels, the user must select the last level before setting the passwords for other levels. A failure to select this particular level first means that the user will not be able to set any passwords and would receive a small pink error message "the password has to be entered".

**Task 3**

*The purpose of this task was to examine the usability of creating and testing an HMI user account. Creating an HMI user is a common practice for adding new operators. In this task, we asked participants to create an HMI user interface/device with a login button, then create a user account to test if the created HMI is working. We expected participants to add a new HMI device to the existing PLC project, and then create an HMI screen with a login mechanism. Afterward, they were to create a user account (i.e., username and password) before testing them on their HMI screen. A successful procedure includes the participant being able to login successfully into the created HMI screen using the details of the newly created user.*

To protect applications against unauthorized operations, users are allowed to set up access protection for data and other functions of the PLC by creating accounts and groups for other users. HMI is a user interface that allows users to manage or control industrial control devices.

- To create a new HMI user account, the administrating user must add a HMI device to one's PLC project, a new interface appears and ask users to select a HMI to connect to the PLC.
- Participants are then asked to create a login button within the HMI screen. They would then need to double click "User administration," to create their individual login.
- Under the HMI project, the administrating user must then select the user tab located at the top right corner.
- Then, they must double click "Add new" under the user tab to add a new username and password for a new user. They will be required to confirm the password before being asked to ratify the changes. The administrating user can create and simulate the HMI to test the new login details.

**Figure 1: Tasks used in our study**

designed to test participants' abilities or how long they took, but to understand *how* they configured the mechanisms. We designed three tasks (explained in detail in Fig. 1). The first task required the participants to connect the PLC to the computer using the TIA portal. The second task focused on configuring the access level control of the PLC, while the third task was aimed at examining the usability of creating and testing an HMI user account. Completing Task 1 first was critical since Tasks 2 and 3 were dependent on the accomplishment of Task 1 (if participants failed to complete Task 1, they would be helped to complete the task). The main plan was to get participants to walk the researchers through the whole configuration process, describing what they were doing and their reasoning while being observed by researchers. At the end of each task, we asked participants about the task, how they found the task and the challenges they experienced when completing the tasks.

We chose a connecting task (PLC to PC) because it is standard for operators to set this up before they can configure anything on the PLC (see Appendix A.5 for the steps). Task 2 and 3 involve common mechanisms (i.e., identification and authentication) for protecting the PLC and the processes that the PLC automates. They are commonly used for enforcing identification, authentication and confidentiality [21]. Lastly, we chose a Siemens PLC because Siemens has the highest share of the market [45]. However, at a high level, the steps to configure some of these security mechanisms are similar with most brands (i.e., Allen Bradley/Rockwell, Siemens, and Mitsubishi); (see Appendix A.4).

*Post Study Interview.* The task-based exercises were followed up by semi-structured interview questions, which aimed at understanding their perception of configuring security mechanisms and thoughts on how the configuration process could be improved. For example, we asked participants which task was challenging to complete and what could be improved to make the configuration task easy to complete. Our interview guide is provided in Appendix A.3.

## 3.3 Study Protocol

All the study sessions were conducted in person by the lead researcher except for two, where the second researcher was present. Participants who agreed to take part in the study were first invited to our lab—at their own convenient time. Before taking part in the study, each participant was given the study information sheet and consent form. The study was carried out in two phases: the task-based exercises and the post study semi-structured interview. After consenting to take part in the study, participants completed the three task-based exercises (Task 1 to Task 3). All three tasks took approximately 55 minutes to complete. For the task exercises, we employed two methods: informal cognitive walk-through and think-aloud approach. A cognitive walk-through is a task-specific approach modeled after the software engineering practice of code walking-through used to examine the usability of a product [50]. It allows the researcher to define a single task or set of tasks that participants have to complete while researchers take notes of their observations. To capture these thought processes as participants completed the task, we adopted a coaching method [41]. The coaching method is a variant of think-aloud technique which allows the researcher to probe, prompt, and encourage participants to describe their actions while completing the tasks. We probed and observed participants to increase the reliability of our data. There is evidence that users sometimes do not report their actual practices during studies [57, 58]. Observation ensured data triangulation which is usually missing in many usability studies. Once the task-based exercises were complete, the lead researcher conducted the final exit interview with each participant. Each interview took an average of 15 minutes. After the interview, each participant completed a short demographics form. At the end of each session, participants were reminded of the study objectives and the withdrawal process. The lead researcher then thanked them for volunteering to take part in the study.

## 3.4 Pilot User Study

Before we finalized our study design, we conducted a pilot study with two participants with knowledge of ICS and configuring PLCs. The goal of this pilot study was to confirm that the task objectives were clear—we would get meaningful results, and the study was not time-consuming and physically demanding. In addition to simplifying the language, the pilot study also helped us to reduce task complexities. For instance, we removed parts of the task that involved using WIFI and connecting to servers. While pilot study participants helped to improve the study, their sessions are not included in the analysis and the results.

## 3.5 Recruitment

Evaluating mechanisms that require specific skills is challenging because it necessitates recruiting skilled individuals who configure these mechanisms as part of their jobs. These individuals may be biased toward what they currently use and might not fairly report challenges due to familiarity and adaptability. In contrast, inexperienced users may struggle with usability due to lack of experience or first-time use but can offer fresh perspectives on the configuration process. To reduce biases and maximize feedback, we recruited both experienced (n=13) and inexperienced participants (n=6). Experienced participants configure PLCs as part of their job or have proficient skills, while inexperienced participants may have some exposure but consider themselves less skilled.

Participants were recruited through professional networks and encouraged to invite colleagues (snowball sampling [17]). A screening questionnaire ensured a balance of demographics and familiarity with PLC brands. Despite efforts to balance gender, most respondents were male, reflecting the male-dominated field of control system engineering [49]. We recruited 19 participants: 16 had configured PLCs before, with 13 considering themselves experienced, and 3 having some experience but not considering themselves skilled. The remaining 3 knew about control systems but had no configuration experience. Participants included professionals from electrical, electronics, mechanical, mechatronics, computer, software, and civil engineering. Most were full-time employees (6) or engineering students (5), with others being contractors, trainees, or retired. Table 2 in Appendix A.1 summarizes the demographics of our study.

## 3.6 Data Analysis

Once the study sessions were transcribed, two researchers independently analyzed the observation notes and interview transcripts using an inductive thematic qualitative approach [7, 12]. The second researcher initially coded the lead researcher's observation notes and two transcripts to create a detailed codebook. The lead researcher then coded one of the same transcripts, refining the codebook with additional low-level codes. Both researchers met to merge similar codes and refine descriptions. After agreeing on the codebook, they proceeded with coding the remaining transcripts. They used Cohen's kappa [13], a statistical measure that evaluates the level of agreement between researchers who classify items into mutually exclusive categories. The inter-coder agreement (i.e., the degree of consistency among two researchers who independently assign codes to a set of qualitative data), measured by Cohen's kappa, was 0.78, indicating substantial agreement [13]. Similar codes were grouped into themes covering various areas. The research team discussed key themes around users' perceptions, suggestions, and challenges in configuring PLC security mechanisms. Any coding disagreements were resolved collaboratively through discussions (arguing to consensus [26]) or by seeking a third opinion if needed.

## 3.7 Threats to Validity and Limitations

Despite our efforts to recruit a balanced sample, this study has several limitations. First, the specific skills required for configuring PLCs limited our participant pool, affecting diversity as all participants identified as males. This may overstate male preferences and overlook other genders' perspectives. Future studies should use more representative samples to investigate gender effects on PLC security usability. Second, our participants, being trained PLC operators, might not have reported all issues due to familiarity and workaround habits. The study involved specific tasks using TIA Portal and Siemens software, introducing potential biases from participants experienced with these tools or challenges for those unfamiliar with them. Despite this, the study offers varied perspectives on PLC security usability. Third, while our study is the first to explore PLC security configuration usability, the results are not generalizable. However, our review of other PLC configurations showed similarities, suggesting that the same challenges might occur with other vendors (See Appendix A.4). Fourth, the order of tasks could have influenced findings, as participants might compare later tasks with earlier ones or experience fatigue, affecting their attention and explanations. Lastly, we recruited 19 participants, which is suitable for qualitative research, reaching saturation by the 13th script and ensuring no new codes emerged by the 19th interview. These limitations do not invalidate our findings; the study provides valuable insights into PLC security configuration usability.

# 4 RESULTS

Our results spans four key usability concepts: communications, navigation, visuals, and features as shown in Figure 2. Our theme *Communication* includes notions around how the interface communicates with participants, whereas *Navigation* theme is more involved with the participants' interactions with the interface while configuring mechanisms. *Visuals*, as a theme, includes concepts that are about what the participants see on the interface, and *Features*, as our last theme, covers interface items needed to complete the configurations. Using these concepts, we discuss the factors that underpin participants' usability challenges while configuring PLC security mechanisms (Section 4.1). We then discuss participants' usability perceptions (Section 4.2), and suggestions that participants shared to improve usability (Section 4.3).

## 4.1 Where The Complexity Lies

We observed several complexities which leads to misconfigurations or pose challenges to users. These observations are not based on participants reports, but what we as researchers found during the study. We discuss these complexities below and show through Figure 3 the three tasks and where complexities lie. Figure 3 also shows which steps during the tasks pose more challenges or where most participants faced challenges.

**Unclear error messages.** When participants configure the access level controls while they are still online, the interface greys out most fields and prevents them from making any changes. There is an error message which informs participants that the local configuration and online file are not identical. The majority of participants got confused by this error message and ended up trying to find ways of syncing both files. We observed similar confusion when participants encountered other error messages.

**Confusing use of familiar words.** Participants were attracted to familiar labels. For instance, while searching for *Access Level Controls*, participants were more likely to click on 'Security Settings' than on 'Protection and Security'. In the TIA portal, 'Security Settings' contains security policies which confused some participants. We observed that using familiar words to mean things with which users were unfamiliar left some users frustrated. The use of familiar labels to mean something else also meant participants took more time searching for the right features and items.

**Complex operations.** Configuring *Complete Protection* in the access level control mechanism was complex and confusing to some participants. To configure this, participants are required to select 'Complete Protection' (the last option on the list) but then set passwords to the top three levels of the access level control mechanisms. This process left some participants confused as to why they had to put passwords on the levels which they had not selected. Some expected to provide a password to just the selected option, not the other levels since *Complete Protection* meant all other levels were protected.

**Lack of guidance.** We observed that the interface does not provide any guidance to users to help them complete the configuration tasks. Also, it does not anticipate possible misconfigurations from users and does not react kindly to the configuration errors when they happen. For instance, when users attempt to configure access level control while online, the interface does not let users know that they needed to go offline to make any changes to the configuration files. It only alerts them about having non-identical configuration files.

**Too Many Clicks.** The number of clicks users have to perform to navigate to the right mechanisms introduces another source of complexity. The interface gives users the ability to navigate to these mechanics through various routes. However, some of these routes are longer and introduce mistakes. For example, to configure access level controls, users are required to navigate to 'Protection and Security,' which for some participants involved too many clicks, especially those who used the tree on the left panel. A high number of clicks meant users were likely to make mistakes or found locating the controls they needed more challenging.

| | Causes of Complexities | Users' Usability Perceptions | Users' Usability Suggestions |
|---|---|---|---|
| Communication | • Unclear error messages<br>• Confusing use of familiar words<br>• Complex operations | • Poor communication<br>• Too much information<br>• Misleading terminology, icons, and nudges<br>• Complex Navigation | • Tell me what I am doing wrong<br>• Reduce information and number of items<br>• Don't just blink tell me more |
| Navigation | • Lack of guidance<br>• To many clicks | • Complex navigation flow<br>• Experience needed<br>• Unfamiliar layout and features | • I need a better flow process<br>• What's next |
| Features | • Lack of confirmation dialog and Push button | • Unfamiliar layout and features | • I want to verify<br>• Help me 'save' |
| Visuals | • Crammed interface<br>• Unusual ordering<br>• Poor signals or unfamiliar cues<br>• Flash LED feature<br>• Grey color scheme | • Windows 98' vibe<br>• Unfamiliar layout and features<br>• Too much information<br>• Misleading terminology, icons and nudges | • Give colors a meaning |

**Figure 2: The key findings from our study can be summarized into four high-level concepts, Communications, Navigation, Visuals, and Features.**

**Lack of confirmation dialog and buttons.** The lack of confirmation messages confused some participants during all the configuration tasks. After making changes, they expected to have a confirmation dialog either to confirm their action or to be notified of the configuration changes. Also, since there are no confirmation messages, we observed that most participants resorted to self-verification of the configurations or parameters. Self-verification prevented some errors, but not all, moreover, it was tedious. Similarly, participants expected a 'save' button to push their changes, especially after setting the IP address. Without a button, most hesitated to continue with the setup. We also noticed that participants who were uncertain about whether their changes were saved, they had closed and reopened the configuration window to confirm their changes.

**Crammed interface.** We observed participants struggle to find features or items they needed to configure the security mechanisms. Due to the number of items, they spent more time searching for relevant items. Most of them would hover over icons to get their description, while others scrolled through all the panes in search of items. Many items and features on the interface made locating specific features difficult, frustrating, and time-consuming.

**Unusual Ordering.** The ordering of the access level control settings confused most participants. When asked to configure complete protection for access level control, most participants attempt to change the first option (top option) on the list, which is *No protection.* The ordering appears to be unnatural considering the nature of mechanism; participants expected the top option to provide complete protection.

**Poor signals or unfamiliar cues.** Participants misinterpreted the signals from the interface during the configurations. They confused the orange/amber colors for mistakes and errors. When a PLC goes online successfully, an amber/orange line appears at the top row. To the majority of participants this implied there were some errors when this happened.

**Flash LED feature.** To verify if the PLC is connected to the PC, the interface allows users to flash the LED on the actual PLC device through a checkbox 'Flash LED' option. However, during the connection tasks, some of the participants did not intuitively use this checkbox to test if the PLC is connected, they directly clicked 'Go online'. This could be due to a number of factors such as: they have not seen the feature, they do not find it necessary, or unsure of what it does.

**Grey Color Scheme.** The use of color grey as the primary background color for the interface confused most of the inexperienced participants, particularly concerning editable fields. There seems to be little to no difference between editable fields and the grey background. Many participants failed to distinguish the differences hence ending up confused as to what they needed to do to complete the tasks. Moreover, concerning greyed out items, we noticed that often participants would attempt to expand the screen in order to have the full view of the screen. Other participants would scroll around or randomly click on the screen, hoping to get some response from the interface.

## 4.2 Participants' Perceptions of Security Usability

Our observation also reflected in the think aloud and interviews; and additional themes were added. Qualitative analysis of data from the think-aloud protocol and interviews revealed that participants' usability perceptions span three main concepts: communication, experience, and visuals.

**Poor Communication.** The majority of the participants highlighted that the interface was not useful at communicating with them during the configuration process. They stated that the interface lacked dialogs, confirmation messages, and clear error messages. They expressed that the lack of confirmation and error messages leads to confusion, explaining that they were never assured about whether they were completing some steps correctly.

*"I think the confusing part for me is going offline to edit stuff and then coming back online"* P16, Inexperienced

The majority of participants reasoned that the error messages from the TIA portal were horrible and not useful during the configuration process.

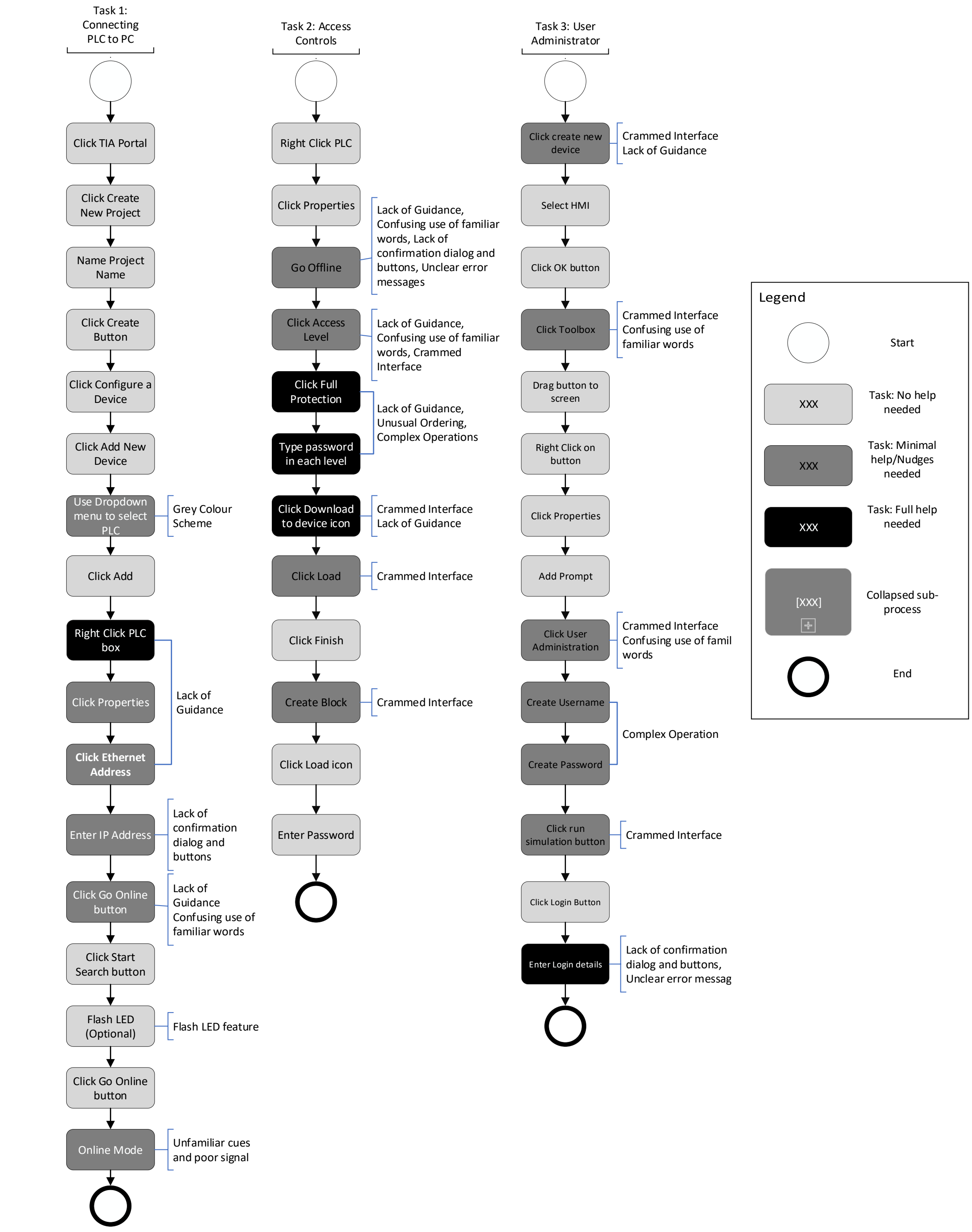

(a) Task 1: Connecting PC to PLC

(b) Task 2: Access Control

(c) Task 3:User Administration

**Figure 3: Mapping between the steps and the complexities that pose various challenges to users while configuring security mechanisms in Siemens PLC.**

*"Oh [curse] is it doing authentication by you ask for whatever you want? Oh [curse], that's horrible"* P8, Inexperienced

Some participants highlighted that the use of the grey color within the interface was confusing at times, especially when it involved a grey text-box to fill or edit. These participants explained that some grey boxes were editable while others were not, so in most cases, they confused the two, either thinking it was not editable when it was or vice versa.

*"I can still click on it, but this one is in grey. But before when I was on grey, it means I couldn't click on it."* P5, Inexperienced

Moreover, when some participants could not edit the disabled grey text boxes, they assumed that it was because they did not have elevated privileges. *"Purely based on not being given admin rights, as such."* P3, Experienced

*"... so, I need to login, I suppose, with another user in order to have elevated privileges."* P9, Experienced

Other participants stated that the interface failed to show that their changes were accepted after updating some particular fields. For instance, after updating the IP address of the PLC, the interface did not signal to them that their changes were updated and saved. *"...Then there's no kind of save or anything so I'm just going to go out of it. And because there's no save, I'm rechecking."* P12, Experienced

**Too much information.** All the inexperienced participants stated that the interface had too much information and it was distracting. *"I think with five PLC you should not need all this out to you. It would distract you and frustrate you by this amount of information."* P11, Experienced

Two of our participants explicitly reported that the interface was not pleasant to use, and it caused some eyestrain after a while. They stated that having to search for items around the screen and the small text caused some stress to their eyes.

*"Is there a way to make the screen bigger? Cause I am sure squinting my eyes is a health and safety problem."* P13, Experienced

**Misleading terminology, icons, and cues.** A point of consensus among participants was how certain terms, icons, and cues were used. The majority reported that the interface was full of 'bad' labels and wordings, which at times led to confusion.

Some participants reported that the TIA portal had misleading cues. They stated that the TIA portal used cues with which they were unfamiliar, for instance, using an orange color to show that the PLC is connected rather than using the green color. They explained that orange or amber usually signified a warning while green meant everything was okay.

*"...There is the orange bar to prove it. Which I think should be a neutral color. I think if there is a warning it would turn orange. Danger is red. Green means its connected. But neutral color should be nothing has changed – no sign or danger."* P14, Inexperienced

They informed us that they were attracted to "security" while setting access level controls because it was more related to what they wanted to configure, but instead, they found that it contained information about security policies. Also, concerning the labeling, some participants cited that some wording was vague and confusing, for example, "Go online," participants explained that it is not clear because it could mean connecting the PLC to the internet, which was not what they wanted to do.

*"Although, I think that it is slightly misleading, if I'm honest, because online you'd think of the internet as opposed to a local connection."* P17, Experienced

**Complex Navigation Flow.** According to our participants, navigation around the interface is complex; finding items is challenging, and the interface lacks directions to help the user. They suggested that the interface has many items, especially menus, which is confusing. Some experienced users stated that, since they do not always configure these mechanisms, they also find locating items challenging because of the number of features on the screen. Others informed us that they expected the interface to help them during the configuration process, for instance, the interface showing them the next step.

*"I would like instructional messages for example the online and offline when doing the online and offline. I was confident when the lights flash, for sure. Because it shows a successful and completed task or connection."* P13, Experienced

Several participants informed us that there are too many "clicks" involved in completing the tasks, while others stated that they did not understand why they sometimes have to double click while completing the task. Some participants suggested that it was difficult to predict where an item would pop-up sometimes items opened at the top other times at the bottom. This made it challenging for them to notice changes in the interface. In addition to the above factors, other participants cited the lack of 'save' button as unusual which made configuring security mechanisms a challenge.

*"Something to save my settings. Or is that already saved?"* P19, Experienced

**Experience needed.** Several participants acknowledged the difficulty of configuring the security mechanisms and expressed the need for training or getting used to the interface. Some suggested that some experience is needed to be able to configure these mechanisms. One experienced participants said: *"When I was just an apprentice – it was [curse] a lot to take in. But after a few years under my belt, it was not hard to do it, but I can't say I understand it completely."* P1, Experienced.

**Unfamiliar Layout and features.** Other participants informed us that the order of the access level control was unfamiliar. They stated that they expected the control that provides the maximum protection to be at the top rather than the bottom of the list since it covered all the levels. Participants also informed us that the interface had many boxes or panes, which made locating certain features challenging.

*"I think initially when you're looking at the interface, there's so many categories, subcategories, items that potentially could lead to [mistakes]. I was a bit overwhelmed at first as to... I think it's possibly more to do with just looking around, looking through the interface and trying to find what you're looking for and the icons does not explain for itself. It can sometimes be mistaken for something else"* P16, Inexperienced

**Windows '98 Vibe.** Some participants, particularly inexperienced

users, suggested that the interface was old, messy and industry like, while experienced users suggested that it was 'ok' for industry standard. Some suggested that the grey color was not approachable, while others explained that too many unnecessary features made locating security features challenging. *"Yeah, it does have a... Windows '98 vibe."* P7, Experienced
*"It is not the color, it is also the whole organisation of everything, like it is so dated, so messy, so industrial product like."* P6, Inexperienced

### 4.3 Usability Suggestions

We now present participants' suggestions on how the usability of PLC security mechanisms configurations can be improved. We sought participants' suggestions to gain insight into their priorities, aligning with the recommendations from previous studies [27, 47, 48, 60, 64] in the field of usable security. These studies emphasize the importance of involving users in system designs. These suggestions have been organized into four themes: (1) communication; (2) navigation; (3) features; and (4) color scheme. We include participants' quotations from the interviews to represent their views and ground the emerged themes.

1. Communication

**Tell me what I am doing wrong.** The majority of the participants would like to have more informative and clear messages. They suggested that error messages should be improved to provide detailed and clear information and at the right time. For example, participants needed to go offline to complete Task 2 or set up Access Management, but the interface does not make this clear. Many participants were left frustrated by this because they did not understand what they were doing wrong. *"I don't care what the interface looks like, I just want it to tell me when I'm going wrong..."* P8, Inexperienced.
The majority of our informants suggested that the interface should help them complete the setup configurations, for instance, instructions on what to do next. *"But I think some changes would be desirable to have such as the instruction messages for the access levels would be useful to have."* P1, Experienced.
Others suggested improving the wording: *"It should say 'Live Mode' or something to say that this PLC is now live."* P13, Experienced.
**Reduce information and number of items.** Some participants suggested that the amount of information on the interface should be reduced, stating that the reduction of information would help navigation. Others suggested reducing items in the interface, such as the number of menus.

*"I think that it can be improved, yes. For example, you can display devices here and reduce the scope. Yeah, but you still need a lot of time to try and get the right answer, I think. It's a lot of information here."* P9, Experienced.

**Don't just blink, tell me more.** Two of the participants recommended that the interface should not just present cues but provide more information. This was about Task 1, where participants were asked to connect the interface to the PLC, when the PLC is successfully connected to the portal, one of the LED lights on the PLC flashes.

*"It should not just blink, but not provide more info..."* P6, Inexperienced.

2. Navigation

**I need a better flow process.** Other participants proposed that the flow of operation should be simplified. One participant further noted that the challenge was not the information but the flow of what they needed to do.

*"I would not simplify the amount of information, but I would develop the 'flow' of how the information is presented."* P11, Experienced.

**What's next!** Some participants, particularly those who struggled with navigation, suggested that the interface should provide some help or direction on what to do when they are stuck. For example, after adding the IP address during Task 1, what users should do next is not made obvious.

*"If I have to improve the TIA Portal, it would be to give me directions and save buttons."* P1, Experienced.

3. Features

**I want to verify.** In relation to the number of concerns over verification, the majority of participants recommended the addition of the confirmation mechanisms to the setup process. For example, after adding a new password to the access level, participants mentioned it would be beneficial to confirm that the password has been updated. *"I am checking my steps again and again, so it's nice to have a confirmation about if I've completed one step in order to proceed to the next one."* P9, Experienced.
**Help me 'save'.** The majority of the participants proposed a 'push' button to allow them to save their changes. They highlighted that majority of the procedures do not include a save or push button; as a result, not making it evident to them that their changes have been saved.

*"I even commented that they should have you know, put an 'apply' or 'whatever' button"* P6, Inexperienced.

4. Color scheme

**Give colors a meaning.** While most of our participants suggested color change, some specifically recommended that error color should be improved, for instance, the color used in cues should change based on the configuration status. They suggest this would help them understand the configuration status better. When participants went online during Task 1, an orange bar/line appeared on the portal followed by a few icons; this led to some participants assuming they had encountered an error which was not the case. This was just a sign they had connected to the PLC, but they interpreted it as a warning sign. *"Color should change based on configurations"* P14, Inexperienced.

## 5 RECOMMENDATIONS

Our findings show that best practices from Human-Computer Interaction (HCI) studies have not been implemented in ICS environments, leading to poor usability of security configuration mechanisms. As a result, usable security in ICS is approximately 20 years behind IT settings, as highlighted by seminal works from 1999 [48, 60]. Our recommendations, based on HCI best practices, address these issues and highlight the importance of usable security. We emphasize that embedded practices (e.g., legacy terminology) and the specific nature of ICS (e.g., devices designed for longevity rather than usability) pose significant challenges. Table 4 maps users' challenges, where these challenges occur most frequently, and potential solutions. Using these factors, we aim to underline the critical need for improving usable security in ICS environments, advocating for the integration of established HCI principles to enhance usability and security effectiveness.

**Addressing interface design issues**. Participants found the wording, colors, and cues used in the interface confusing. For example, while configuring the connection between the PLC and the PC, participants misunderstood terms such as "Go Online," interpreting it to mean connecting to the internet. Additionally, when asked to connect the PLC using an IP address, most participants searched for "network configurations" instead. The use of orange color to indicate a successful connection led to misinterpretations, with participants thinking it signified a warning. Participants also found searching for a particular feature time-consuming. By following current interface design patterns, many of these problems can be addressed. Tidwell [54] suggests that a simple user interface is necessary when users face challenges finding items. Displaying the most important features upfront and allowing users to access hidden items with a single, simple gesture can save time.

> **Recommendations:**
> 1. Use clear and modern interface terminologies and cues, e.g., the orange color should be used to show a warning instead of a successful connection.
> 2. Adopt modern and commonly used interface design patterns. For example, use breadcrumbs and labels to show users the path from where they started to their current page. This can also include showing users only features relevant for the task at hand.

**Confirmatory feedback**. Most participants found it difficult to proceed with configurations without confirming their changes. This is counter-intuitive to their mental models, which expect a button or a confirmation dialog to approve changes before proceeding, as seen in more user-friendly interfaces. Wash [59] on mental models demonstrates that people often apply mental models from other settings to the current situation. In our study, participants preferred to manually confirm their changes by closing and re-opening the configuration window, which was time-consuming. Without save buttons and confirmation dialogs, it is challenging for users to know whether their changes have been applied. Implementing these features would also improve navigation and communication of errors—users would understand what to do next and what they are doing wrong. This aligns with usability principles that emphasize feedback and user control, making interfaces more intuitive.

> **Recommendation:**
> 1. Add confirmation dialogs to help users confirm changes.
> 2. Improve primary actions such as saving changes by including and making a 'save' button standout.

**Reduce confusion over navigation path**. Our findings show that participants click many times to find the right mechanisms. Because of this some participants mix and confuse navigation paths during configurations. We also recommend that the number of steps to configure some mechanisms such as access level control should be reduced. We contend that this would reduce complexity (reduce confusion over navigation path) and effort required to configure the mechanisms.

> **Recommendation:** Reduce cognitive load and complexity by reducing the number of steps to complete tasks. This can include navigation tabs, breadcrumbs, and progressive disclosures. For example, paths can be broken down into sections, or non-essential information and items could be hidden from users.

**Simplify configuration process**. In addition to decreasing the navigation steps to reduce complexity, we also suggest simplifying the configuration process by providing users with cues during the configuration of security mechanisms. In our study, many participants explained that the interface lacked guidance and direction. Most participants struggled with configuring the access level control mechanisms because of the way they were presented. Using a more straightforward mechanism, where users can quickly tick and provide a password for the desired access level, would reduce the complexity of configuring access levels. We also observed participants spending much time searching for the right items; some used tooltips to find the right icons. Providing such help would reduce the time taken to configure and verify the settings and allow PLC operators to complete other responsibilities.

> **Recommendation:** Provide suitable cues to help users complete tasks. Use design patterns such as wizards; break the task into dependable sub-tasks.

**Testing mechanisms**. Participants struggled to test their configurations; they spent a lot of the time trying to figure out how to test whether their configurations were successful. We posit that this is due to the lack of instructions to verify configurations and unusual testing mechanisms. We, therefore, recommend that the testing mechanisms and instructions should be clear to users. They should also be visible to avoid complexity.

> **Recommendation:** Introduce testing mechanisms with clear instructions. Users could be shown the new configuration before they continue.

| Task | Challenges | Description | Hotspots | Recommendation |
|---|---|---|---|---|
| Task 1 | Grey Color Scheme | Participants mistaking unavailable features as available due to grey color scheme | – Use Drop-down menu to select the right PLC | • Use clear and modern interface terminologies and cues. |
| | Lack of Guidance | Participants struggling to find features and/or properties. | – Right Click on PLC<br>– Configure the Ethernet Address<br>– Click Go Online | • Provide suitable cues to help users complete tasks.<br>• Use clear and modern interface terminologies and cues.<br>• Provide updated documentation as soon as it is available. |
| | | Participants not knowing how to proceed. | – Right Click on the PLC | • Reduce cognitive load and complexity by reducing the number of steps to complete tasks.<br>• Provide suitable cues to help users complete tasks. |
| | Lack of confirmation dialog and buttons | Lack of usable methods to verify setup | – Configure the Ethernet Address | • Improve primary actions such as saving changes. Include and make a 'save' button standout.<br>• Add confirmation dialogs to help users confirm changes. |
| | Confusing use of familiar words | Participants struggled with the use of some terminologies. | – Click "Go Online" button | • Use clear and modern interface terminologies and cues. |
| | Flash LED feature | Lack of instruction on how to test connectivity. | – Flash LED feature. | • Introduce testing mechanisms with clear instructions. |
| | Unfamiliar cues and poor signal | Participants struggled to interpret signals and errors. | – Online Mode | • Use clear and modern interface terminologies and cues.<br>• Provide suitable cues to help users complete tasks. |
| Task 2 | Lack of Guidance and Lack of confirmation dialog and buttons | Participants not understanding that they needed to go offline before continuing with the configuration. | – Go Offline | • Provide suitable cues to help users complete tasks. |
| | | Participants not knowing that they should download the configuration file to the device | – Click "Download to device" | • Provide suitable cues to help users complete tasks.<br>• Introduce testing mechanisms with clear instructions.<br>• Provide concise instructions on how a task can be completed. |
| | Crammed interface | Participants struggling to find items on the screen, e.g., Access Level Control mechanisms. | – Click Access Level<br>– Click "Download to device"<br>– Click Load<br>– Create Block | • Provide suitable cues to help users complete tasks.<br>• Reduce cognitive load and complexity by reducing the number of steps to complete tasks.<br>• Use clear and modern interface terminologies and cues. |
| | Unusual ordering, Complex operations | Participants found the access level control mechanism layout confusing. | – Click Full Protection<br>– Type password in each level | • Use clear and modern interface terminologies and cues.<br>• Provide concise instructions on how a task can be completed.<br>• Provide a better help mechanisms such an inline help box. |
| Task 3 | Lack of guidance, Crammed interface | Participants struggling to find items and features on the screen. | – Create Username<br>– Create Password<br>– Click User Administration<br>– Click "run simulation" | • Provide suitable cues to help users complete tasks. |
| | Complex operation | Participants struggling to distinguish the difference between the process of creating individual user accounts and creating groups. | – Click create new device<br>– Click Toolbox<br>– Click User Administration<br>– Create Username & Password. | • Reduce cognitive load and complexity by reducing the number of steps to complete tasks.<br>• Provide concise instructions on how a task can be completed.<br>• Provide a better help mechanisms such an inline help box. |
| | Lack of confirmation dialog and buttons | Participants failing to notice that they have logged in | – Enter Login details | • Provide suitable cues to help users complete tasks.<br>• Add confirmation dialogs to help users confirm changes. |

**Figure 4: *Challenges, Hotspots and Potential Solutions*: Mapping between challenges, hot spots and the ways in which they can be addressed.**

**Help mechanisms**. During all the tasks, some participants used their devices to search for solutions when faced with configuration challenges. Only two participants attempted to use the inbuilt help mechanism but decided against it after realizing how it was organized and worked. Providing a suitable help mechanism within the application would help operators stay within the secure environment and not risk using external resources that could leave the plant vulnerable (e.g., connecting to the internet to search for solutions). Moreover, providing clear instructions on how to perform certain tasks can significantly aid users in configuring complex settings, such as Access Level Controls. This aligns with usability best practices, emphasizing the importance of integrated support and clear guidance within the application to enhance both user experience and security.

**Recommendations:**

1. Provide a better help mechanisms such an inline help box.
2. Provide concise instructions on how a task can be completed.
3. Provide updated documentation as soon as it is available.
4. Provide post report of changed configurations to help reduce misconfigurations.

# 6 DISCUSSION

## Lack of "Product Usability"

Our analysis indicated that PLCs are generally designed with a lack of usability considerations but for reliability and endurance [39]. TIA Portal was designed to configure PLCs out in the field when there were no powerful computers [35]. The interface resembles what operators will see on the HMI, crammed with a long, complex navigation flow. There is also an issue of terminology; the terminology used in the ICS field has not evolved the same way as in the IT space, "going online" in an ICS environment does not mean connecting to the internet. These differences cause much confusion on operators who are both IT and ICS devices users. We posit that solving usability issues will not work by simply updating the TIA portal. ICS equipment run critical infrastructures, which must provide robust safety features and real-time properties that operators may not be interested in tempering with. Moreover, simply upgrading the portal to provide usable security may disregard the expertise of the operators who configure these PLCs and make systems work. Security, in this case, will get in the way of engineers, and it may be costly to train all operators on how to use the new portal. Overall, our study and findings reveals the interplay between the design constraints and the operators' usability perception of configuring PLCs. It highlights the different aspects (absent in IT environments) that designers of PLC need to consider to help improve security in PLCs. Our findings highlight the importance of usable security in legacy systems and offer solutions to these challenges. Further research is needed to assess usable security in other ICS devices, such as HMIs, to develop a more refined set of requirements.

## Experience vs Usability

Participants' usability suggestions imply that experience (or training) plays a vital role in the perception of usability. As we suggested in Section 3.5, experienced users tend to develop mitigations strategies over time and may ignore some usability challenges they face. For example, experienced participants in our study did not find any issues with not having a save button and the number of items on the screen. They did not suggest any improvements regarding them. This study presents the first evidence that operators and their role may play a vital role in bringing usable security to PLCs—changing the interface may significantly impact operators' everyday tasks—leading to more misconfigurations. Studies should consider that PLC operators are also IT users, HCI principles and patterns more established in IT settings affect them in their everyday life outside the work environment. Research should explore how these differences influence PLC usability, focusing on operators' expectations, common mistakes, and recovery methods.

# CONCLUSION

We illustrated through this study that the security of legacy systems, especially that of PLCs, lags behind other sectors. Our findings show that usable security of such systems matters and requires special attention from security researchers and practitioners. Particularly, there is a need to consider the specific design and deployment contexts in this regard—systems often lack the computational capability and graphical interfaces that are the norm in IT environments while operators need to contend with terminology and configuration processes that may be counter intuitive to what they typically encounter and "learn" in IT settings. Future studies should, therefore, aim to understand these nuances with regards to terminologies, idiosyncrasies of the industrial environment as well as constraints such as inability to update the systems due to their longstanding nature and need for uninterrupted operation and up-time. Designs for usable security will need to consider the interplay between such constraints and best practice HCI principles and guidelines. Our study and design recommendations are a stepping stone in this regard.

# A APPENDIX

## A.1 Demographics Table

**Table 2: Demographics of the participants table. A total of 19 participants took part in our study. Sixteen (16) participants identified as Male, while three (3) were as female. Regarding experience in configuring PLCs, thirteen (13) considered themselves experienced while six (6) considered themselves as less experienced. In the last row, we show the number of participants that needed help to complete the three tasks.**

| Category | Experienced | Inexperienced |
|---|---|---|
| **Age** | | |
| 21 - 25 | 0 | 3 |
| 26 - 30 | 2 | 2 |
| 31 - 35 | 3 | 1 |
| 36 - 40 | 2 | 0 |
| 41 - 45 | 2 | 0 |
| 46 - 50 | 3 | 0 |
| 50 + | 1 | 0 |
| **Employment status** | | |
| Unemployed/Retired | 1 | 0 |
| Full time | 7 | 2 |
| Self employed (Contractors) | 2 | 0 |
| Student | 2 | 3 |
| Apprentice | 1 | 1 |
| **Industrial Sector** | | |
| Computer Eng. | 0 | 2 |
| Software Eng. | 0 | 2 |
| Electrical/Electronics Eng | 4 | 2 |
| Research and Development | 3 | 0 |
| Mechatronics/Robotics Eng. | 4 | 0 |
| Civil/Structural Eng. | 2 | 0 |
| **Configured Control System** | | |
| Yes | 11 | 2 |
| No | 2 | 4 |
| **Experience Brand** | | |
| Siemens | 6 | 1 |
| Schneider Electrics | 3 | 1 |
| Rochwell Automation | 4 | 1 |
| ABB | 3 | 0 |
| Omoron | 2 | 0 |
| Others | 13 | 6 |
| **Gender** | | |
| Male | 12 | 4 |
| Female | 1 | 2 |
| **Help Needed** | | |
| Task 1 | 2 | 4 |
| Task 2 | 4 | 5 |
| Task 3 | 5 | 5 |

## A.2 Practical Study Script

We present our study guide and interview questions below.

*Introduction:* I have assumed you have read the participants information sheet? If not, don't worry I will give you a bit more detailed brief. So in front of you is a Siemens PLC. This study is about completing a practical series of PLCs configuration tasks. You will be guided if you found yourself stuck; and this is not in any form a test of your abilities or capabilities. We asked you to use a think aloud protocol whilst completing this study. So please, just talk about what you see on the screen and what your thoughts are. There are four practical studies and a short interview after. This should take roughly 40 to 50 minutes of your time. Do you have any questions before we start?

*Task 1: Configuring PLC to PC.* To get started, please open TIA portal and create a new project. You can call it whatever you wish.

You have created a new project with the right PLC. But it needs to be connected. The first task is to link this PLC to the computer you are on. I have written down the IP address in front of you. You may use this or configure it as a direct Ethernet cable. Your choice.

- Where would you first look?
- Why would you look in there?
- What words are you searching for?
- Would you use the help function?
- How do you know its connected?
- How would you test it?
- (If participant did not Flash LED) Would you flashing the LED lights make you confident that the PLC and PC are connected?
- That concludes this study, how did you find it? and what did you find difficult and easy?

*Task 2: Configuring Access Levels.* In the second task you are going to have to protect this PLC using a password. I need you to go to access level and set up four passwords for the four accesses. These are Read access, HMI access, no protection and complete protection. Where would you first look?

I would need you to setup a password for each level. Please remember your passwords as this will be used to test if the configuration was a success. I can make note on your passwords or you may use the suggestions on the piece of paper in front of you. Your choice.

- Where would you get the access level?
- Why do you think its greyed out?
- Is it because you are online? Why do you not think that is the case?
- Can you tell me what the pink message box is saying?
- Have you tried double clicking the box?
- Why do you think its not allowing you to enter the password?
- Could you now try adding the passwords please?
- How would you test that the configuration has been a success?
- That concludes this study, how did you find it? and what did you find difficult and easy?

*Task 3:HMI simulation/User Administration.* Final part of task 2. You are going to create an HMI simulation. You want to create a login button with its function as you seen on this screen, and I'll run you through the other things as we go along. So you need to add a HMI first.

- So, how would you create a new device or create a new HMI?
- So you have configured the HMI to link the PLC. Now I want you to get a login button as you can see on this snapshot. I also want you to give its function as shown here. Where would you add the button to your HMI?
- So where would you add its function?
- Right, so you added a login button to your HMI, can you create yourself a user login and password to test this?
- Do you think its because you are not a user administrator?

- Whats the difference between user group and individual?
- Could you start the simulation?
- Where do you think is the start simulation?
- That concludes this study, how did you find it? and what did you find difficult and easy?

## A.3 Post Study Interview Script

This is the last part of the study, so in this short interview I want you to reflect on your whole experience in this study.

- Can you tell me what your perception of the user interface was?
- How did you find completing the tasks?
- What task was easy and why?
- And what are the challenges you found when trying to complete the task?
- Why did you find it difficult? (would you say the interface was difficult due to a lack of dialogue?)
- How would you improve the interface?
- What would you change to the PLC configuration design?
- Did you understand the information flow between the PLC and PC? ?
- How do you feel about this overall study and do you have any questions?

## A.4 PLC comparison flow diagrams

Figure 5 to 7 shows the tasks steps and what participants found challenging.

### A.5 PLC Interface - steps to Task 1 (Connection PLC to PC)

To connect a Siemens S7-1200 PLC to a computer, the user must load the TIA portal. If no project exists, they should click 'create new project.' After creating the project, the user needs to configure the device by adding it and selecting the correct PLC type and version from the drop-down menus.

- The next step involves them to right-click on the added PLC and then properties (Step 1 and 2, as shown in Fig. 8).
- On the resulting screen, under the 'General' tab, the participant is then required to locate and click on 'ethernet addresses' on the inside left pane (Step 3, Fig. 8).
- The user is then asked to provide the method of connection, ethernet cable, or provide a valid IP address (Step 4, indicated in Fig. 8).
- After this, the user must then click 'Go online' (Step 5, Fig. 8). To confirm if the PLC is connected, the participant must search for the PLC by clicking on 'start search' (Step 6, 8) .
- Once the PLC is found, the option to flash the LED lights on a PLC is made available. By checking the 'Flash LED' box will flash the LEDs on the actual PLC device. This is optional (Step 7, Fig. 8).
- Once completed, the user would click 'Go Online' (Step 8, Fig 8) The 'Go Online' means connecting to the PLC so one can upload logic or send commands.

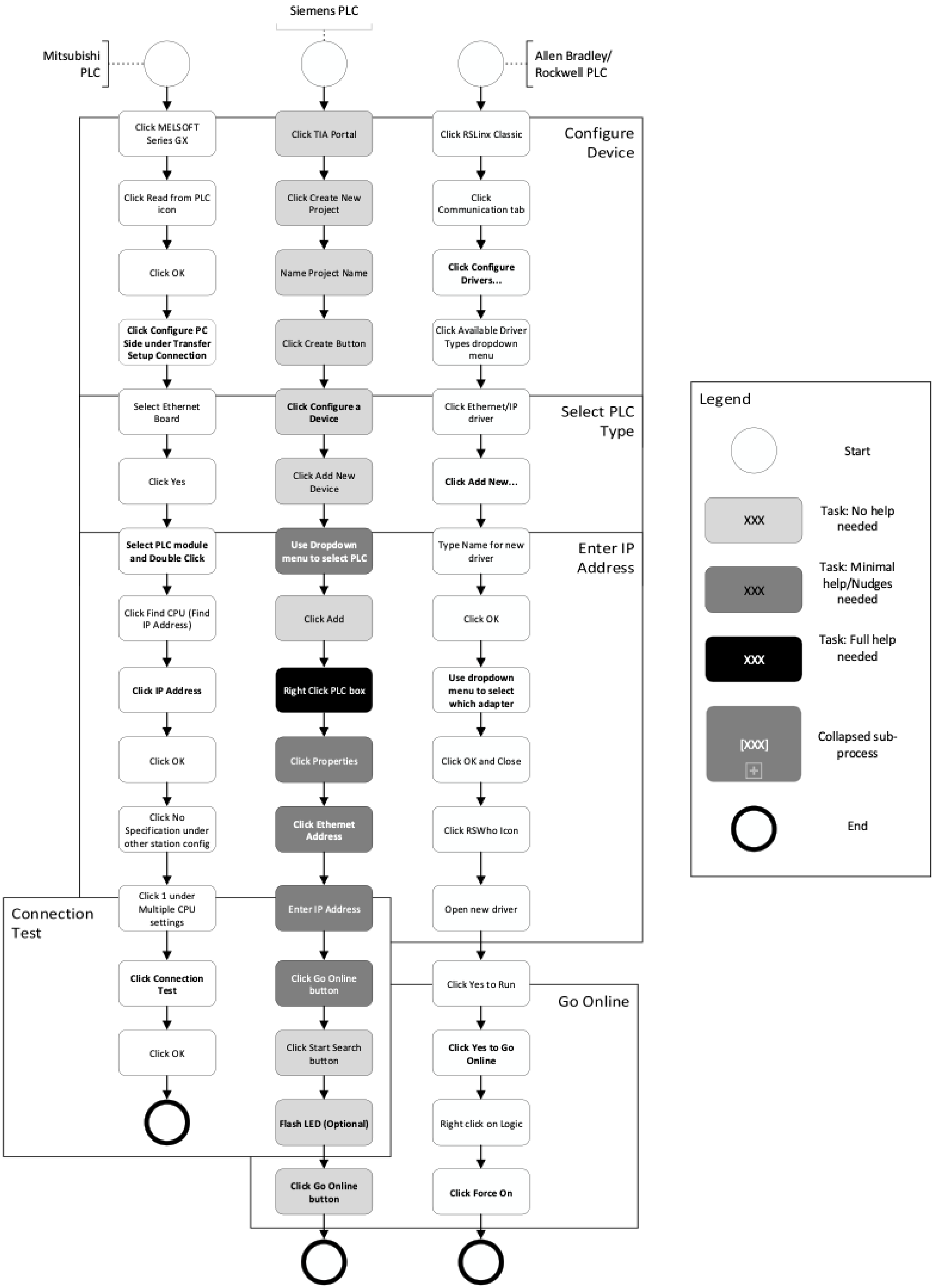

**Figure 5: Task 1: Connecting PLC to PC steps comparison**

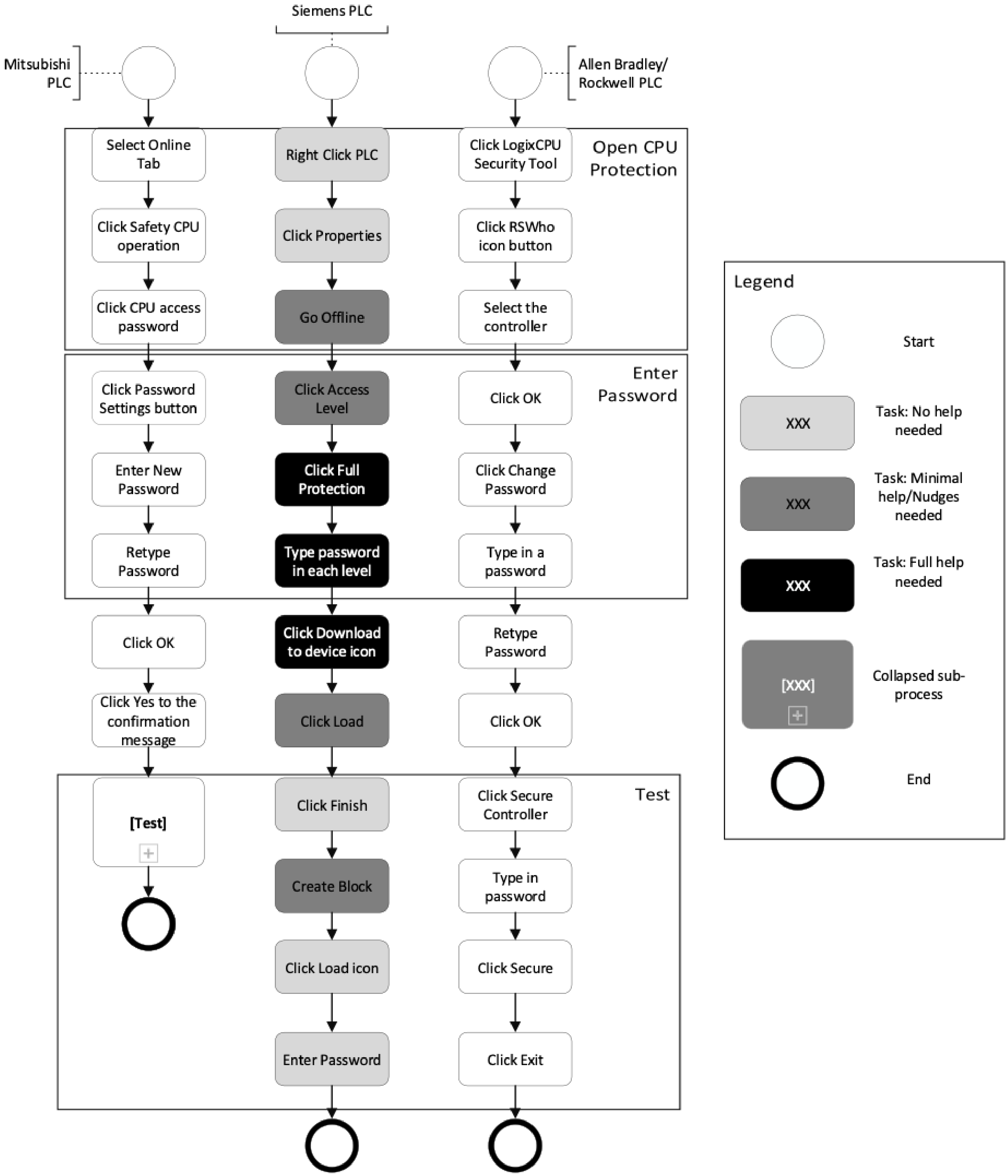

**Figure 6: Task 2: Access Levels steps comparison**

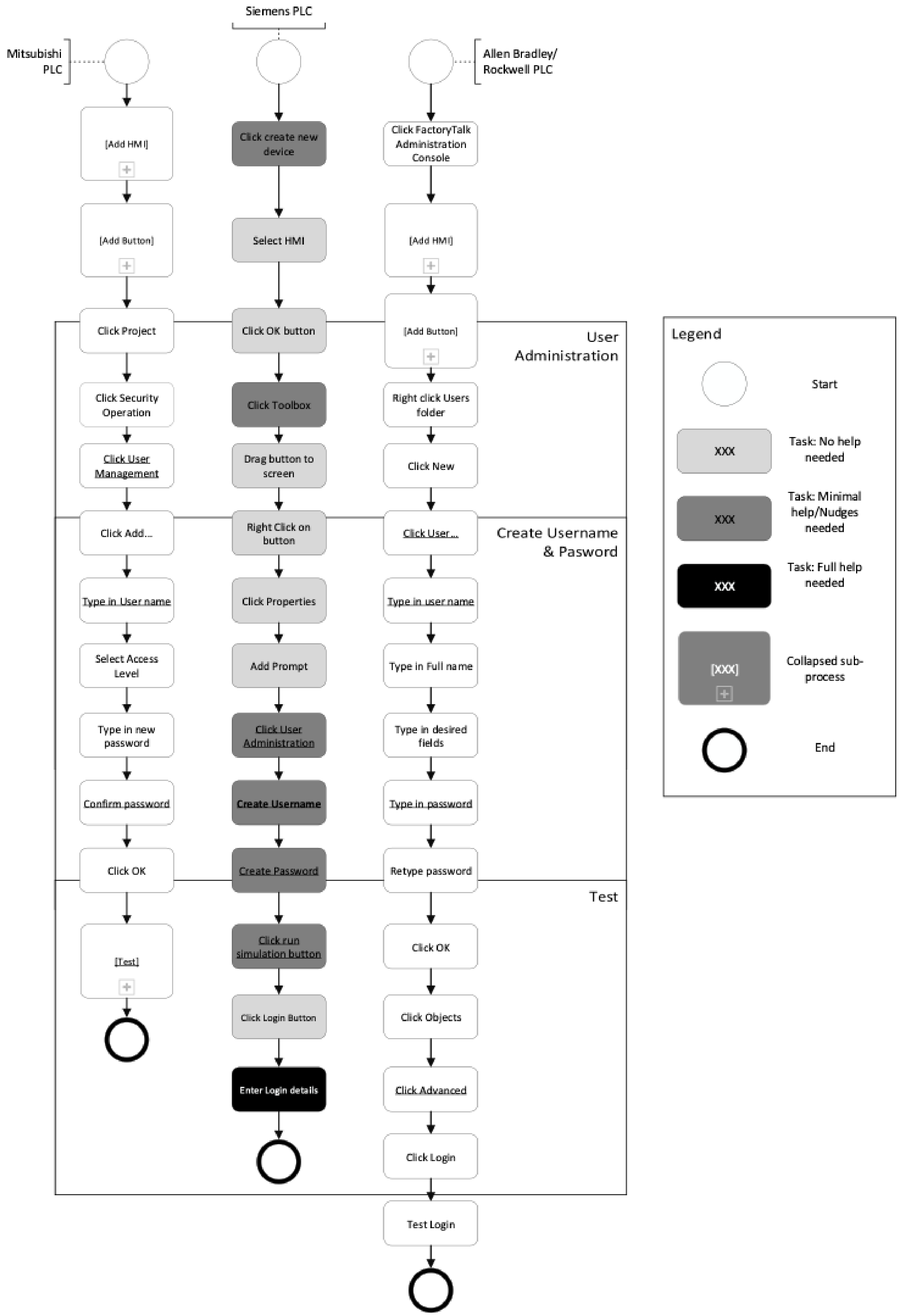

**Figure 7: Task 3: User Administration steps comparison**

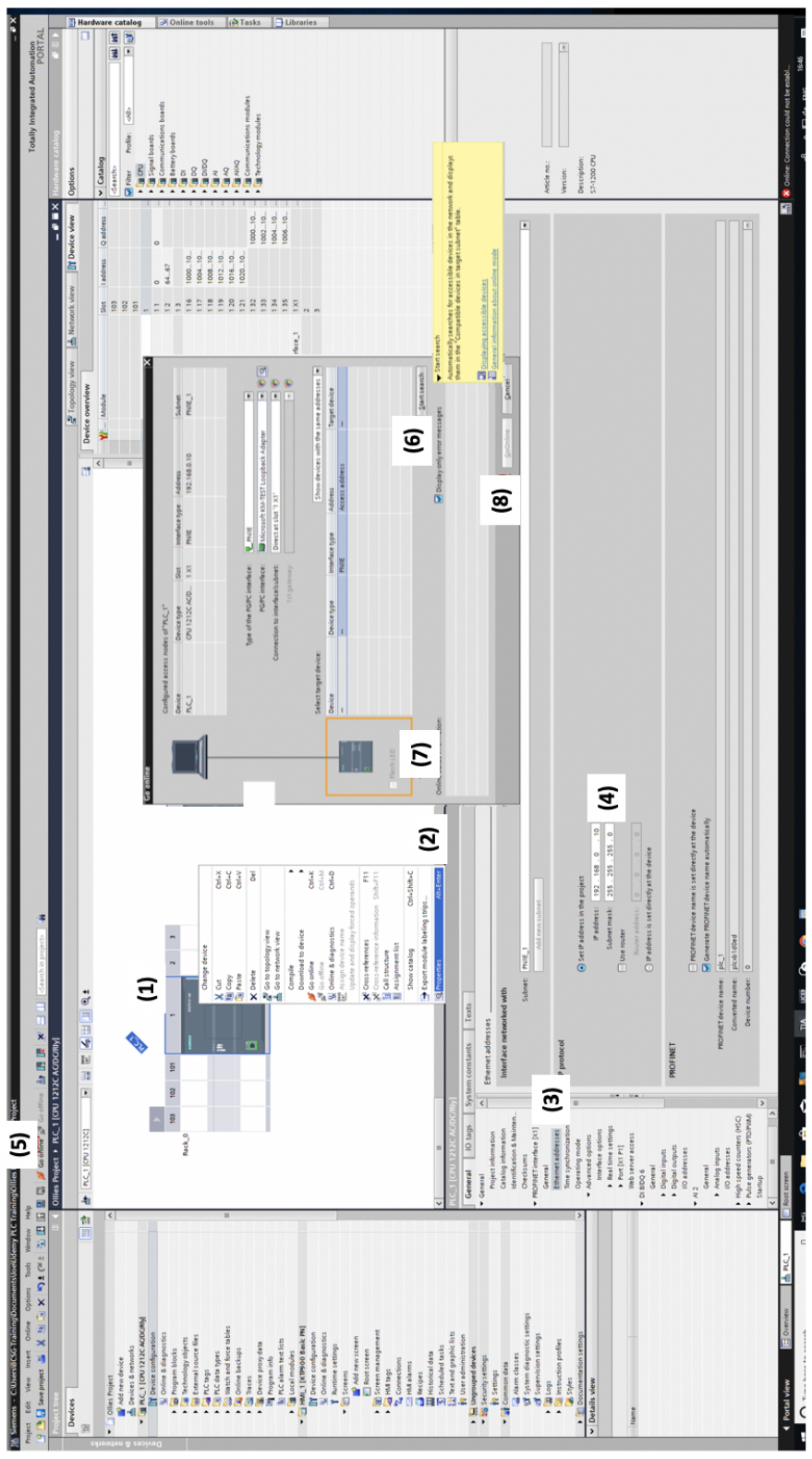

**Figure 8: TIA Portal Interface: Connecting PLC to PC Steps (1) to (8)**